\documentstyle[prb,aps,multicol,epsf]{revtex}
\twocolumn
\begin{document}
\title{
Conductance of
a quantum point contact in the presence of spin-orbit interaction
}
\author{Shi-Liang Zhu$^{1,2}$ , 
Z. D. Wang$^{1,3}$ \thanks{To whom correspondence should be addressed.
Email address: zwang@hkucc.hku.hk}, and Lian Hu$^{2}$
}
\address{ 
$^{1}$Department of Physics, University of Hong Kong, Pokfulam Road,
Hong Kong, China\\
$^{2}$Department of Physics, South China Normal University,
Guangzhou 510631, China\\
$^{3}$ Department of Material Science and Engineering, University of Science and Technology of China, Hefei, China
}

\address{\mbox{}}
\date{J. Appl. Phys. 91, 6545 (2002)}
\address{\parbox{14cm}{\rm \mbox{}\mbox{}
A recursive Green's function technique
is developed to calculate the spin-dependent conductance
in mesoscopic structures. Using this technique,
we study the spin-dependent electronic transport 
of quantum point contacts in the presence of the
Rashba spin-orbit interaction.
We observed that some
oscillations in the `quantized' conductance are induced by
the spin-orbit interaction, and indicated that the oscillations
may stem from the spin-orbit coupling associated multiple reflections.
It is also indicated that
the $0.7$ structure of the conductance
observed in mesoscopic experiments
would not stem from the spin-orbit interaction.
}}
\address{\mbox{}}
\address{\parbox{14cm}{\rm PACS numbers:  
71.70.Ej, 73.23.-b, 73.20.Dx}}
\maketitle   

\narrowtext
\newpage

\section{Introduction}

The interaction between the electron spin and its orbital motion,
commonly refered to as the spin-orbit (SO) interaction or SO coupling,
has been known for a long time.
Recently the effect of SO-interaction on mesoscopic
transport phenomena and the quantum Hall
effect has attracted growing
interest.~\cite{Meir,Loss,Aronov,Zhou,Zhu,Morpurgo,Moroz,Bulgakov}
Although the interaction 
magnitude is small compared to the Fermi energy, it may have
significant impact
on electronic transport, particularly in mesoscopic systems where
quantum interference is extremely important.
Using the transfer matrix method, Meir {\sl et al.}~\cite{Meir}
showed that the SO-interaction in one-dimensional (1D) non-interacting
disordered rings induces an effective spin-dependent magnetic flux, and
then any spin-dependent transport quantity
can be expressed in terms of the same quantity in
the absence of the SO scattering but with an effective magnetic flux.
The adiabatic Berry phase induced by the SO-interaction
and its effect on the electronic transport
were studied extensively by Loss {\sl et al.},~\cite{Loss}
Aronov and Lyanda-Geller.~\cite{Aronov}
Persistent currents in mesoscopic rings
induced by the SO-interaction was addressed in Ref.4.
Promisingly, as observed by Morpurgo {\sl et al}.,~\cite{Morpurgo}
the geometric
phase induced by the SO-interaction~\cite{Meir,Loss,Aronov,Zhou,Zhu}
could induce the splitting of the main peak
in the ensemble average Fourier spectrum.
Also interestingly, lifting of the spin degeneracy by
the Rashba SO-interaction\cite{Rashba}
was reported experimentally in two-dimensional (2D) electron systems
for different semiconductor structures.~\cite{Luo}
Lommer {\sl et al.}
pointed out that the Rashba mechanism becomes
dominant in a narrow gap semiconductor system.~\cite{Lommer}
The spin splitting energy at the
Fermi energy in the absence of magnetic field 
is $\Delta=2\alpha k_F$
(with $\alpha$ as a SO coupling constant and
$k_F$ as the Fermi wave vector), which
is about $1.0\sim 5.0\ meV$ in typical semiconductor
materials.~\cite{Luo}

On the other hand, the observation of the universal~\cite{Wees,Wharam}
and the nonuniversal~\cite{Thomas} quantizations of the
conductance
in a quasi-1D constrication is also remarkable. 
In a clean quasi-1D constrication, if the mean free
path is much longer than the effective channel length, the
conductance is quantized in unit of
$2G_0$ with $G_0=e^2/h$ at zero magnetic field,
referred to as the universal quantization, 
where the factor of $2$
aries from the electron spin degeneracy.~\cite{Wees,Wharam}
Recently, Thomas and co-workers 
found that, in addition to
the above quantized conductance plateau, there is also
a structure at $0.7(2G_0)$. This so-called
$0.7$ structure (nonuniversal or fractional
conductance quantization)
appears to be related to spontaneous lifting
of spin degeneracy in the 1D constrication,
but the origin of it
remains as an open question.~\cite{Thomas}
Since the $0.7$ structure may be understood as
a zero-field spin splitting with an estimated energy of
$\Delta\sim 1 meV$,~\cite{Thomas}
which is the same order as
the spin splitting energy induced by the SO-interaction,
it is nature to ask an important question:
Is the spin polarization induced by the SO coupling
responsible for the $0.7$ structure?
To answer this question, we 
study the effect of SO-interaction on the
electronic transport through
quantum point contacts (QPC) in this paper.
We find that the $0.7$ structure is
unlikely to stem from the SO-interaction. However,
some interesting oscillations in the `quantized' conductance
may be induced by the SO-interaction,
and may be experimentally observable.
Moreover, the spin-dependent transport properties
are of current interest in both fundamental physics
and applied spin electronics.
It is quit intriguing to develop a method
to calculate the spin-dependent
transport in mesoscopic systems.
Also in this paper, generalizing a method
established for spin-independent cases~\cite{Ando,Sanvito}
we have done such a job.  It is worth pointing
out that another theoretical approach\cite{Cuevas} may
be generalized to calculate the spin-dependent conductance.

The paper is organized as follows. In Sec.II A,
we derive the 2D tight-binding Hamiltonian with the Rashba SO coupling.
In Sec.II B, the recursive Green's function
technique is presented
to calculate the spin-dependent transmission coefficient.
In Sec.III the conductance is 
calculated numerically in a model QPC in the presence of SO coupling and/or
magnetic field. The finite temperature effect is also studied.
This paper ends with a brief summary.

\section{Model and calculation method}

\subsection{2D tight-binding Hamiltonian with the
Rashba SO coupling}

\begin{figure}
\label{fig1}
\epsfxsize=7.5cm
\epsfbox{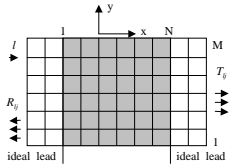}
\vspace{-4.0cm}
\caption{Schematic diagram of the system.
Electrons are injected
from the left lead.}
\end{figure}

In a magnetic field, the 2D Hamiltonian for electrons
(of effective mass $m^*$ and charge $e$)
with the Rashba SO-interaction reads~\cite{Rashba}
\begin{equation}
\label{Hamiltonian}
H=\frac{{\bf \Pi}_x^2+{\bf \Pi}_y^2}{2m^*}
+\frac{\alpha}{\hbar}
(\hat{\sigma}_x {\bf \Pi}_y-\hat{\sigma}_y {\bf \Pi}_x)+U(x,y)
-\frac{g\mu_B}{2}
{\stackrel{\hat{\rightarrow}}{\sigma}}\cdot {\bf B},
\end{equation}
where ${\bf \Pi}={\bf p}-\frac{e}{c}{\bf A}$ is the
canonical momentum, $U(x,y)$ represents
the spin-independent potential, ${\bf A}$ is
the electromagnetic gauge potential with
${\bf B}=\nabla \times {\bf A}$ relating it to the magnetic fields,
and $g$, $\mu_B$,${\stackrel{\hat{\rightarrow}}{\sigma}}$
are the $g$ factor, Bohr
magneton, Pauli matrices respectively.
$\alpha$ is the Rashba SO-interaction constant which
is determined from an
effective electric field along $z$ direction given by the
form of the confining potential in the absence of an
inversion center.
Now we choose a discrete square lattice, on which
points are located at $x=na$ and $y=ma$, with
$n,m$ as integers and $a$
as the lattice constant.
In terms of the quaternion,
the one-electron
tight-binding Hamiltonian with the Rashba SO coupling
can now be parameterized as
\begin{eqnarray}
\nonumber
H&=&\sum\limits_{nm}\varepsilon_{nm}|nm\rangle\langle nm|
-\sum\limits_{nm}(V_{nm,n-1m}|nm\rangle\langle n-1m|\\
&+&V_{nm,nm-1}|nm\rangle\langle nm-1|+H.C.),
\label{2DHamiltonian}
\end{eqnarray}
where $|nm\rangle$ is a two-component orthonormal set in
the lattice sites $(n,m)$;
$\varepsilon_{nm} = (U_{nm}+4t)\tau_0-i\gamma\tau_3$ with 
$U_{nm}=U(x=na,y=ma)$, $t=\hbar^2/2m^*a^2$,
$\gamma=g\mu_B B_z/2$; 
$V_{nm,n-1m} = V_x exp(-ieA_x a/\hbar)$ and
$V_{nm,nm-1} = V_y exp(-ieA_y a/\hbar)$
with  $V_{x}= t\tau_0-\alpha^{\prime}\tau_{2}$,
$V_{y}= t\tau_0+\alpha^{\prime}\tau_{1}$
and $\alpha^{\prime}=\alpha/2a$.
The quaternion basis $\{\tau_i\}$ is defined by
the $2 \times 2$ unit matrix $\tau_0$ and
$\tau_l=-i\hat{\sigma}_l\ (l=1,2,3)$ with $\hat{\sigma}_l$ as
the Pauli matrices.
$A_x$ or $A_y$ is evaluated at the middle point between sites $(n,m)$
and $(n+1,m)$ or $(n,m+1)$.

\subsection{Model and the recursive Green's function technique}

To describe spin-dependent transport properties through a specific
mesoscopic structure,
we consider a 2D structure composed of three different
regions ( Fig.1).
The shadowed central zone is a mesoscopic structure 
where the SO coupling and/or a homogeneous
perpendicular magnetic field are present.
To simplify scattering boundary conditions, we
assume that the structure is connected to both sides with
semi-infinite ideal leads without
SO-interaction and magnetic field.
In our lattice model atoms are 
at the sites of
a square lattice $(na,\ ma)$
( with $a$ to be set equal to unit, $m=1,2,\cdots,M$.)
in the shadowed zone and two ideal leads.
As a result the system is
described by a one-electron tight-binding Hamiltonian 
similar to Eq.(\ref{2DHamiltonian}).
For a homogeneous magnetic field
${\bf B}=(0,\ 0,\ B)$ perpendicular to the 2D-plane
(${\bf A}=(-yB,\ 0,\ 0)$ in the Landau gauge),
we can write the Hamiltonian for the shadowed central zone
as 
\begin{equation}
\label{Hamiltonian-c}
H_c=\sum\limits_{n=1}^N|n)H_{n}(n|
+\sum\limits_{n=1}^{N-1} [|n)H_{n,n+1}(n+1|+H.C.],
\end{equation}
where $|n)$ is the set of $2M$ ket
vectors belonging to the $n$th cell, and
\begin{equation}
\label{n-Hamiltonian}
H_n\equiv H_{n,n}=\left (
\begin{array}{lllll}
\varepsilon_{n1}  & -V_y^+  & 0      & \cdots & 0\\
-V_y    & \varepsilon_{n2}  & -V_y^+  & \cdots & 0\\
0      & -V_y    & \varepsilon_{n3}  & \cdots & 0\\
\vdots & \vdots & \vdots & \ddots & \vdots \\
0      &   0    & 0      & \cdots & \varepsilon_{nM}
\end{array}
\right ),
\end{equation}
\begin{equation}
\label{Hn,n+1}
(H_{n,n+1})_{pp'}=-V_x^+ e^{-i2\pi\beta(p-\frac{M+1}{2})}
\delta_{pp'}\ \ (p,p'=1,\cdots,M).
\end{equation}
Here  $\beta=Ba^2/\phi_0$ with
$\phi_0=hc/e$ as the magnetic flux quantum.
The forms of Hamiltonian for the two ideal leads are the same as
Eq.(\ref{Hamiltonian-c}), but with different summing regions
($-\infty <n<1$ at the left lead
and $N<n<\infty$ at the right lead) and different values of
the parameters:
$V_x=V_y=t\tau_0,\ \ \varepsilon_{nm}=4t\tau_0,\ \ B=0$.
Moreover, it seems acceptable to assume
that the coupling between the ideal leads
and the mesoscopic structure
takes simply the form $H_{0,1}=H_{N,N+1}=-t\delta_{pp'}$.~\cite{Datta2}

Following the recursive Green's function technique developed for
some spin-independent
systems,~\cite{Ando,Sanvito,Lee}
we may generalize the method to the nontrivial spin-dependent cases.
The spin-dependent
transmission  $t_{lj}^{\sigma\sigma'}$
for the incident channel $(l,\sigma)$
and out-going channel $(j,\sigma')$
can be found readily (see the appendix A) .
At a finite temperature $T$,
the conductance through a 2D mesoscopic structure
is given by the Landauer-B\"{u}ttiker formula~\cite{Landauer}
\begin{equation}
\label{Landauer}
G=-G_0\sum\limits_{lj,\sigma\sigma'}
\int_0^\infty T^{\sigma\sigma'}_{lj}
\frac{\partial f(E,T)}{\partial E}dE,
\end{equation}
where $T^{\sigma\sigma'}_{lj}=|t^{\sigma\sigma'}_{lj}|^2$
is determined from Eq.(\ref{coefficient}),
and $f(E,T)=[1+exp(E-E_f)/k_BT]^{-1}$
is the Fermi-Dirac distribution function
with $k_B$ as the Boltzmann constant.
Obviously, Eq (\ref{Landauer}) reduces to
$G=G_0\sum_{\{lj,\sigma\sigma'\}}
T^{\sigma\sigma'}_{lj}$ at zero temperature.

In the following, we focus on
a QPC structure with the B\"{u}ttiker saddle-point potential
\begin{eqnarray}
\nonumber
U_{nm}=V_0 &-& \frac{1}{2}m^*\omega_x^2 a^2 [n-(N+1)/2]^2\\
&+&\frac{1}{2}m^*\omega_y^2 a^2 [m-(M+1)/2]^2,
\label{potential}
\end{eqnarray}
which is a practical candidate to describe a real
QPC.~\cite{Buttiker} $\omega_{x,y}$ indicate the strength of the lateral confinement.
Note that 
the well-pronounced quantized plateaus occur
if $\omega_y\geq \omega_x$.

Before the end of this section, it is worth pointing
out that the recursive Green's function 
method described
in this section can be straightforwardly generalized to
3D mesoscopic structures, and thus is quite useful in studying the
spin-dependent transport of many different mesoscopic structures which
are of current interest, such as the devices
with quantum Hall effect, giant magnetoresistance,
tunnel magnetoresistence.

\section{Numerical results}

We now calculate numerically  
the conductance of the QPC in the presence of
SO-interaction and/or a magnetic field.
We here present the results
for the system of $M=7$, $N=15$ \cite{Sizes}
and $t=1$.

\subsection{The effects of SO interaction}

It is well known that the conductance through QPC is quantized
with the unit of $2G_0$ as a function of the saddle energy
$V_0$. The conductance quantization is well explained by
Landauer-B\"{u}ttiker formula of quantum ballistic transport:
for a nonmagnetic QPC,
the spin-up and spin-down electrons make the
same contribution and then the unit of the conductance quantization is
$2G_0$. However, if the Rashba
SO-interaction is present in the QPC,
the spin degeneracy of conductance electrons
would be broken. Whether 
the quantization of the conductance is affected by the
SO-interaction in a meaningful sense?

\begin{figure}
\label{fig2}
\epsfxsize=7.5cm
\epsfbox{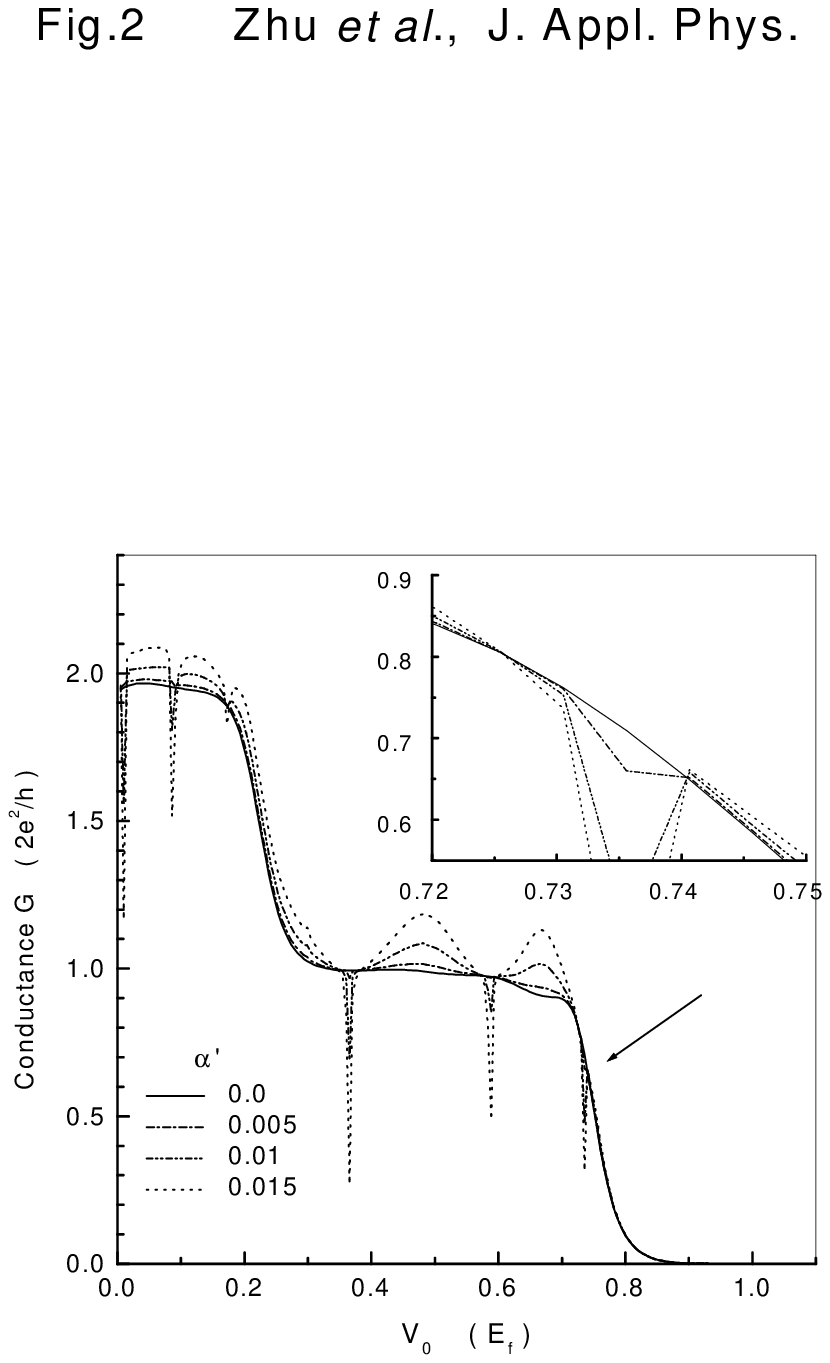}
\vspace{-3.0cm}
\caption{The conductance $G$ versus the saddle energy 
$V_0$ ( in unit of $E_f$)
in the presence of SO-interaction for
$\beta=k_B T=0$, $E_f=1.21$, $\omega_y=3\omega_x$.
The inset shows an enlargement
of the curve around the arrow.}
\end{figure}

Figure 2  shows
the conductance $G$ as a function of 
$V_0$ for the SO parameters $\alpha^{\prime}=0,$ $0.005$,
$0.01$, and $0.015$. Other parameters are $E_f=1.21$, $\omega_y=3\omega_x$.
Two effects of the SO-interaction
on the conductance are seen.
(i) The SO-interaction induces some oscillations 
on the plateaus, and the stronger the SO-interaction is
the larger the oscillation amplitude is. Remarkably, some sharp dips exist
on the oscillation for the strong SO coupling,
for example, the height of dip is almost $2G_0$
for $\alpha'\sim0.015$.
(ii) A tiny plateau exists near $0.7(2G_0)$ for a weak SO interaction
$\alpha\sim0.005$ ( We will address the temperature effect on
this tine plateau later).
But it becomes a sharp dip for the strong
SO-interaction (see the inset of Fig.2).
The conductance oscillations
as well as the sharp dips
in the plateaus
may be observable in future mesoscopic
experiments.

We now attempt to understand the above numerical results heuristically.
The conductance oscillations appear to stem from the multiple reflections
with the SO coupling. 
By taking into account the multiple reflections, 
the conductance of the QPC may be rewritten as~\cite{Wees}
\begin{equation}
\label{reflection}
G=G_0\sum\limits_{j\sigma}\frac{(1-R_{j\sigma})^2}{1-2R_{j\sigma} cos(2k_{j\sigma} L_x)+R_{j\sigma}^2},
\end{equation}
where the reflection probability $R_{j\sigma}$ for channel $(j,\sigma)$ is given by
\begin{equation}
\label{R}
R_{j\sigma}=\left( \frac{v^L_{j}-v^c_{j\sigma}}{v^L_{j}+v^c_{j\sigma}}\right) ^2,
\end{equation}
with $v^L_{j}$ ( $v^c_{j\sigma}$) as the value of  velocity in the leads ( scattering zone).
$v^c_{j\sigma}=\partial E^c_{j\sigma}/\hbar\partial k$ is determined by the dispersion relation
for the electron states in the QPC, which is approximately written as 
\begin{equation}
\label{dispersion0}
E^c_{j}\approx\hbar^2 k_j^2/2m^*+(j-1/2)\hbar\omega_y+V_0
\end{equation}
(independent on $\sigma$) in the absence of the SO coupling,\cite{Wees} but as 
\begin{equation}
\label{dispersion}
E^c_{j\sigma}\approx\hbar^2 k_{j\sigma}^2/2m^*+\sigma\alpha|k_{j\sigma}|+(j-1/2)\hbar\omega_y+V_0
\end{equation}
in the presence of SO coupling. Here we only consider the energy spectrum 
at $x= 0$ since the transport properties for
the B\"{u}ttiker saddle-point potential
depend mainly on the narrowest part.
Note that the quantized plateaus are well-pronounced in the absence of the SO coupling 
(as shown in Fig.2) since $v^c_j=\hbar^2 k_j/m^*\sim v^L_j$ and thus $R_{j}$ for a propagating model in the B\"{u}ttiker saddle-point potential
is negligible. However, 
the additional $R_{j\sigma}$ may be induced from the
SO coupling because as seen from Eq.(\ref{dispersion}) $v^c_{j\sigma}$ in 
the presence of SO coupling is different from that in the absence of SO coupling.
Therefore, the oscillations may stem from the SO coupling since
Eq (\ref{reflection}) predicts the transmission oscillations, provided that $R_{j\sigma}$ induced by 
the SO coupling
are not too small.~\cite{small}
Moreover, we can understand from Eqs. (\ref{reflection}) and (\ref{R}) that a dip in the conductance may appear for a relatively larger $R_{j\sigma}$ when
 an electron in the channel $(j,\sigma)$ is in a propagating state
($k_{j\sigma}>0$) with $v^c_{j\sigma}\sim 0$. 
It is interesting to note from Eq.(\ref{dispersion}) that $v^c_{j-}=0$  
for $\tilde{k}_j= m^*\alpha/\hbar^2$, which corresponds the minimum of $E^c_{j\sigma}$
as a function of momentum. When $V_0$ varies in the scale of magnitude
$\sim \alpha \tilde{k}_j$, $v_j^c\sim 0$ for the same $v_j^L$-channel as in the
absence of SO coupling.
This clearly shows that a larger $R_{j\sigma}$ may appear
in a narrow region in the presence of the SO coupling,
leading to a dip. Our numerical results in Fig.2 agree qualitatively with the above arguments.

We now look into whether the interesting behaviors seen in Fig.2 are sensitive
to the set of parameters.
Firstly, we plot Fig.3(a) and (b)  for two other different $\omega_y/\omega_x$ 
( the other parameters are the same as 
those in Fig.2). The plateaus are not well pronounced
for $\omega_y=1.5\omega_x$ even if $\alpha^\prime=0$(Fig.3(a)).
In this case, the SO-induced oscillations are not so distinctive with the
conductance curve for $\alpha^\prime=0$, but the sharp dips
still appear in the conductance.
For $\omega_y=6\omega_x$ ( Fig.3(b)),
the width of the plateaus is
widened. Comparing the results in Fig.2, Fig.3(a), and (b),
we found that the well-pronounced plateaus
occur for larger ratio $\omega_y/\omega_x$,
which is due to the B\"{u}ttiker saddle-point potential.
We also noticed that the effects induced by the SO coupling are similar to 
those in Fig.2.
Secondly, we plot   Fig.3(c) and (d) for two other different Fermi energies
(the other parameters are the same as 
those in Fig.2).  It is seen that the main effects induced by the SO coupling
are still similar to those in Fig.2. 
Therefore, we may conclude that the essential feature induced by the SO coupling
in Fig.2 is not quite sensitive to the Fermi energy 
and the ratio of $\omega_y/\omega_x$, but is indeed sensitive to the SO coupling parameter
$\alpha'$.

\begin{figure}
\label{fig3}
\epsfxsize=7.5cm
\epsfbox{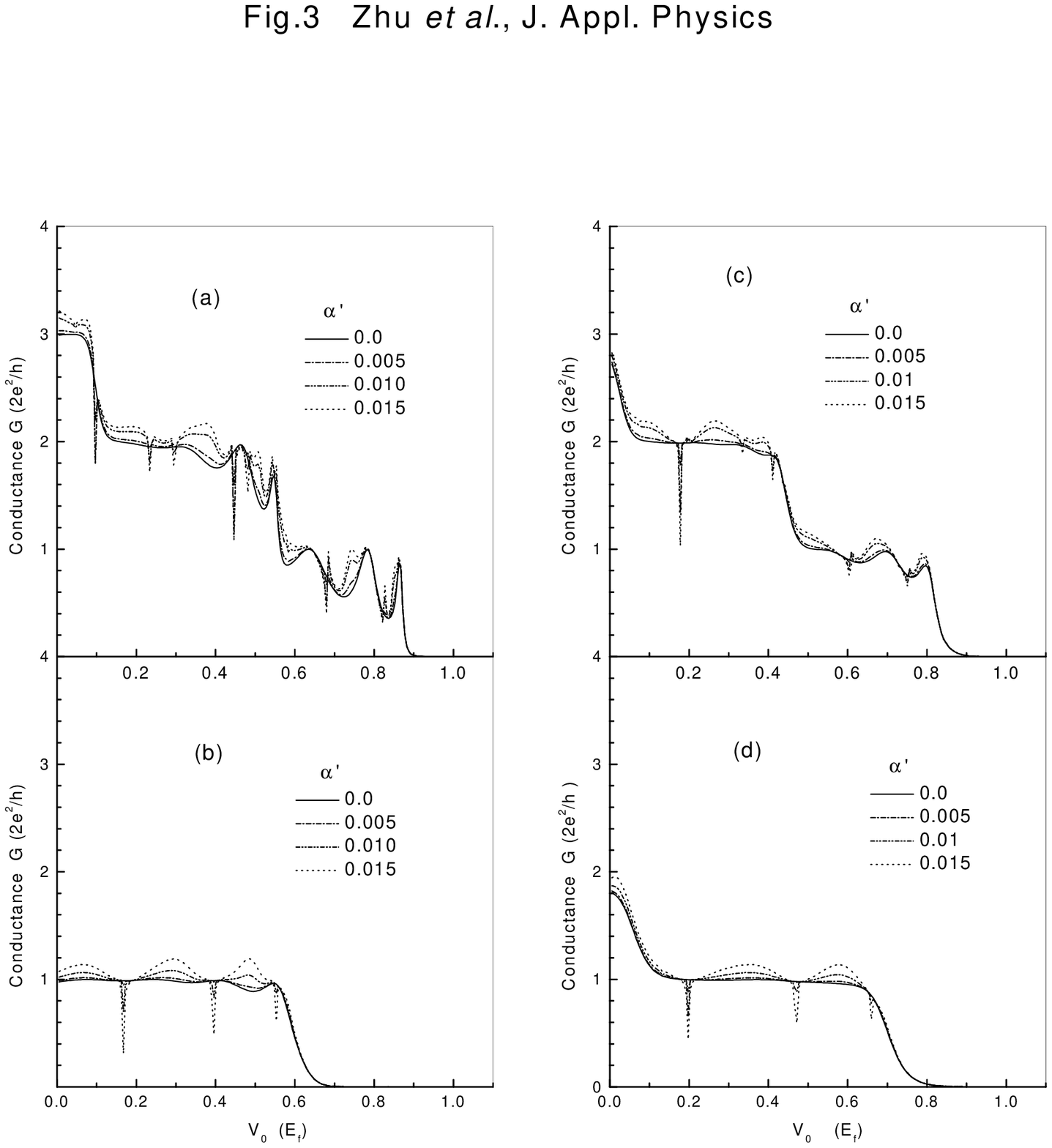}
\vspace{-3.0cm}
\caption{The conductance $G$ versus the saddle energy 
$V_0$ 
in the presence of SO-interaction for
$\beta=k_B T=0$. 
(a) $E_f=1.21$, $\omega_y=1.5\omega_x$;
(b) $E_f=1.21$, $\omega_y=6\omega_x$;
(c) $E_f=2.0$, $\omega_y=3\omega_x$; 
(d) $E_f=1.0$, $\omega_y=3\omega_x$.}
\end{figure}

The temperature dependence of the conductance as a function of $V_0$
can be seen from Figs.4(a) and (b). Firstly, in certain temperature regions,
the oscillation induced by the SO-interaction
disappears when the temperature increases, meanwhile the
quality of the quantized plateaus
is improved. However the
quantized plateaus are destroyed when
temperature increases further.
The mechanism for the destruction
of the quantized plateaus by the finite temperature
is energy averaging.\cite{Wees}
Secondly, the tine plateau and the dip near $0.7(2G_0)$
are destroyed by energy averaging before
the integer plateaus disappear.
However the $0.7$ structure observed by
Thomas {\sl et al.} is still observable even at a
temperature when all the integer quantized plateaus
disappear.\cite{Thomas}
This obvious different temperature effect on the quantization
conductance suggests that the $0.7$ structure
should not stem from the SO-interaction.
Finally, the quantized plateaus disappear at
$k_B T\sim 0.1$ for $\alpha^{\prime}=0.005$, while
the plateaus is still obvious at this temperature
for $\alpha^{\prime}=0.01$ (the plateau disappears at a higher
temperature $k_B T\sim 0.12$).

\begin{figure}
\label{fig4}
\epsfxsize=7.5cm
\epsfbox{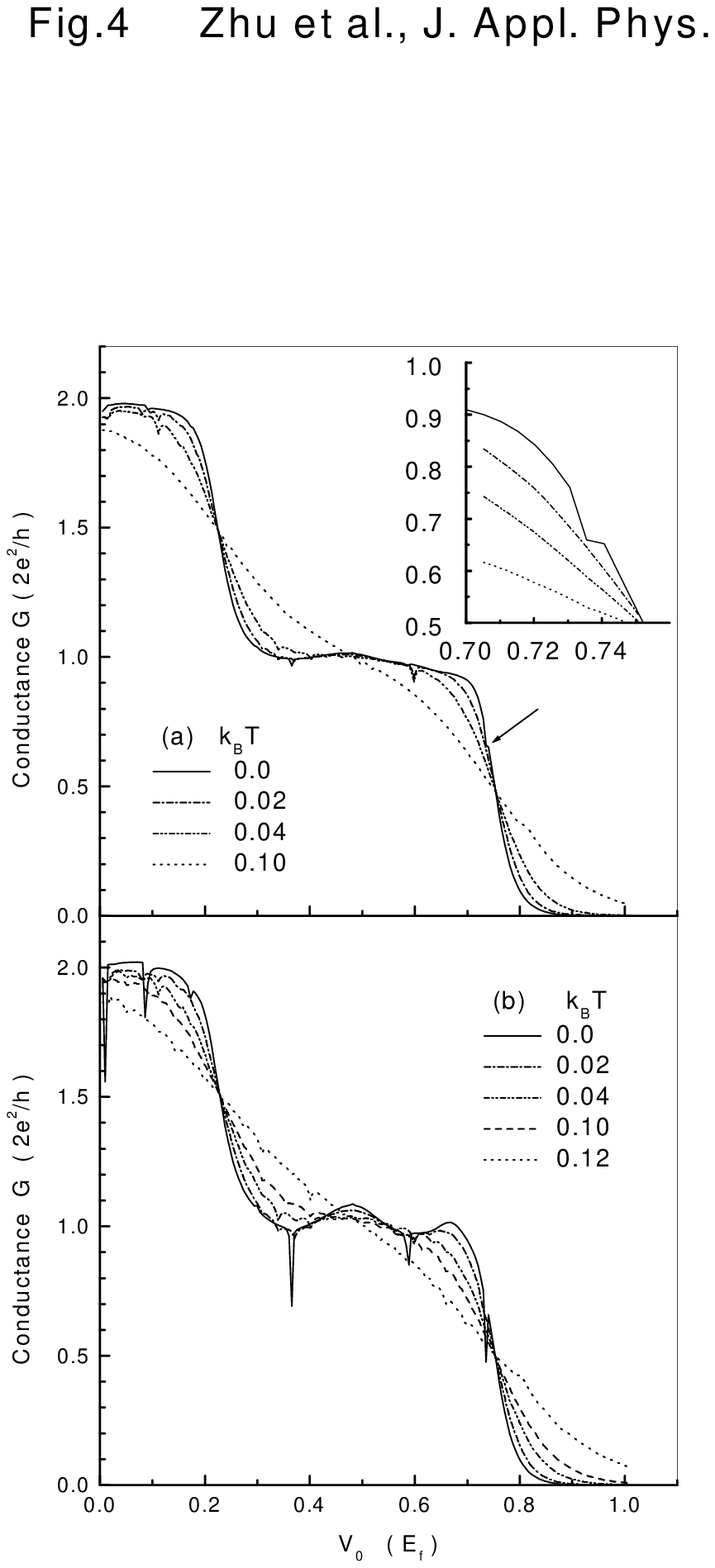}
\vspace{-3.0cm}
\caption{The conductance $G$ versus $V_0$
at finite temperatures for different SO-interactions:
(a) $\alpha^{\prime}=0.005$, with 
the inset showing an enlargement
of the curve around the arrow;
(b) $\alpha^{\prime}=0.01$. Other parameters are $E_f=1.21$, $\omega_y=3\omega_x$.}
\end{figure}

\subsection{The effects of magnetic fields}

We now discuss the effect of a perpendicular magnetic
field on the conductance quantization:
the Zeeman effect and the Peierls phase factor.
In many theoretical calculations, the Zeeman effect
is usually ignored for simplicity. Here we
consider both the Peierls phase and the Zeeman effect.
Figures.5(a) and (b) show the field-dependent conductance
as a function of $V_0$ for 
$\alpha^{\prime}=0,\ 0.01$, respectively.
From Fig.5, one of the characteristic features of
QPC's in a magnetic field is that
conductance steps begin to
appear at odd integer multiples of $G_0$ at $\beta\geq 0.09$,
due to
the lifting of the spin degeneracy by the Zeeman effect.
Another feature is that the width of the
plateaus is widened when compared to the $\beta=0$ case.
On the other hand, the number of effective
subbands is decreased when the magnetic field is applied, and
actually it is proportional to $1/\beta$.\cite{Wees}
Moreover, figure 5(b) shows that the oscillation as well as
the sharp dips are destroyed at high fields.
Thus the quality of the quantization is improved when
the magnetic field is applied.
Certainly, the effect of the SO-interaction is suppressed
by a high magnetic field.

\begin{figure}
\label{fig5}
\epsfxsize=7.5cm
\epsfbox{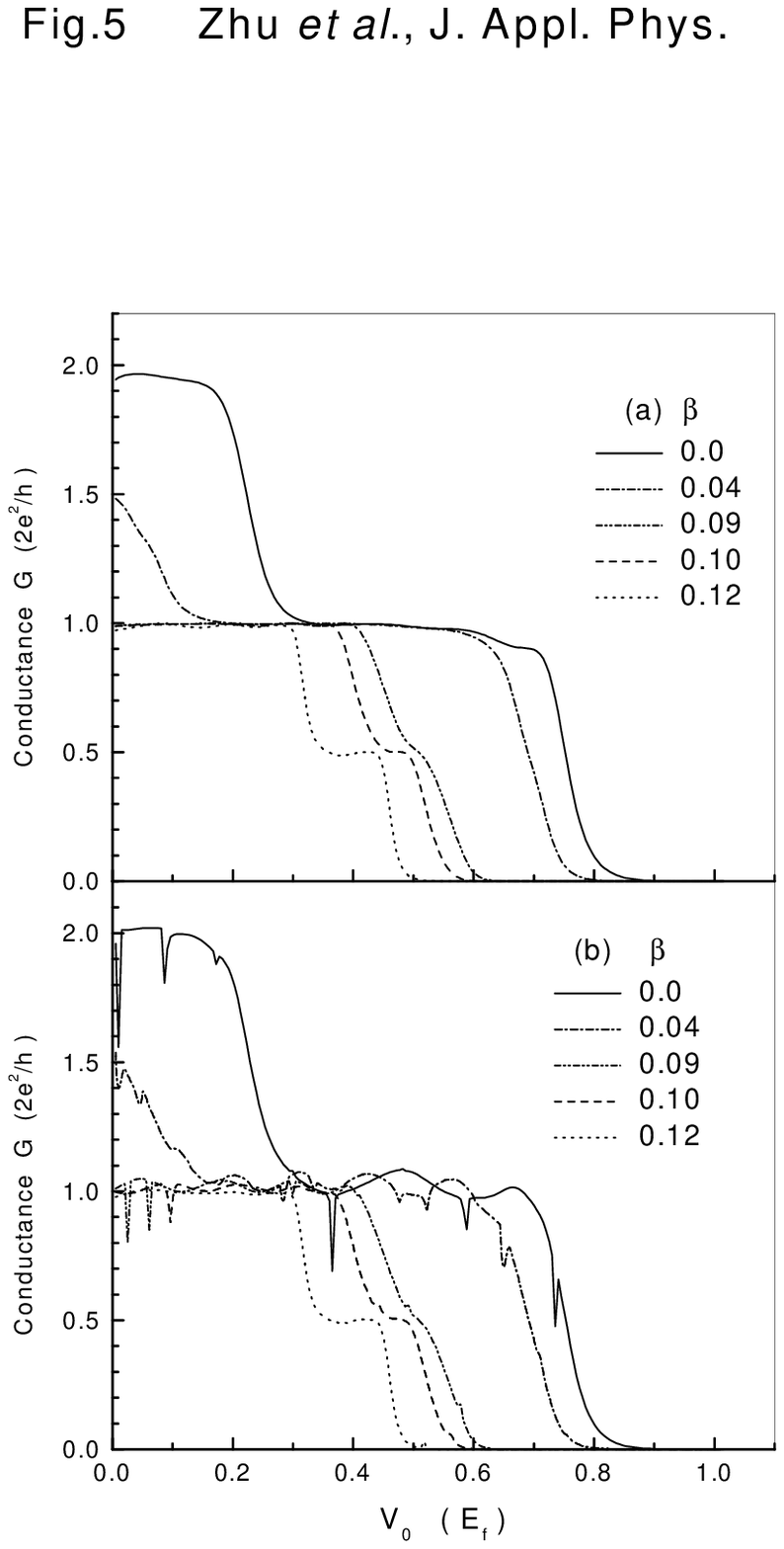}
\vspace{-3.0cm}
\caption{The conductance $G$ versus $V_0$
in the presence of magnetic field.
(a) $\alpha^{\prime}=0,\ k_B T=0$; (b)
$\alpha^{\prime}=0.01,\ k_B T=0$. Other parameters are $E_f=1.21$, $\omega_y=3\omega_x$.}
\end{figure}

\section{Conclusions and discussions}

The SO-interaction is equivalent to a momentum-dependent
effective magnetic field.\cite{Meir,Aronov,Zhou}
However, comparing with a genuine magnetic field,
there exist some
essential differences
in the energy spectra of electrons induced
by an effective field or a genuine field,~\cite{Moroz}
which induce several significant different effects on
the conductance.
Firstly, spin split energy $\Delta$
induced by the SO-interaction
at $|k_i=0|$
is zero, while the spectrum induced by a genuine field
is determined by the nonzero Zeeman energy $g\mu B_z$.
This is the reason why
the conductance steps do not 
appear at odd integer multiples of
$G_0$ in the presence of SO-interaction but
zero magnetic field, noting the appearance of this kind of
quantized plateaus is the essential feature induced by
a genuine field.
Secondly, in the presence of SO coupling,
there exist bumps ( a nonmonotonic portion)
in the spectrum as a function
of transverse momentum,
while the spectrum 
is a monotonous function of the momentum
in the presence of a genuine magnetic field.
Therefore, the reflections at the scattering region
in the presence of SO coupling may induce by 
the negative velocity for $k>0$.\cite{Wees} It provides
us another understanding of the transmission oscillations.
Finally, the SO-interaction can remove the spin
degeneracy in the band structure but could not lead
to an overall spin polarization.\cite{Moroz}
Consequently, the $0.7$ structure is unlikely
induced by the SO-interaction. On the other
hand, the results addressed in the presence of magnetic
field may be restricted within weak field since
the non-interacting electron model
used in the recursive Green's function is invalid in a
very strong magnetic field.

We are now concerned with the possible experimental
test of the above SO-induced properties.
In unit of $t$, $\alpha'=m^* \alpha a/\hbar^2$,
which is  estimated to be $(0.54\sim 5.4)\times 10^{-3}$ for
some typical  semiconductors
[ here we choose the parameters of
InAs, where $a\sim 0.608\ nm$,
$\alpha\sim (1.0\sim 10.0)\times 10^{-10}\ ev\ cm$,
and $m^*\sim 0.067\ m_e$ with $m_e$ as the mass of
a free electron.~\cite{Luo}].
Thus the oscillations
induced by SO coupling may be experimentally observable,
depending crucially on the value of $\alpha'$.

Finally, we wish to make a few remarks 
on why the SO-induced effects on the quantized conductance have not
been observed in experiments so far.  Firstly, the experimental observation on
the SO-induced phenomenon would crucially depend on the value of $\alpha^\prime$, but 
$\alpha^\prime$ in some typical semiconductor systems 
may be relatively small based on the above estimation. Thus 
it is helpful to increase the coupling coefficient 
in experiments. 
 As indicated by
Heida {\sl et al.},~\cite{Luo}
two methods may be used to increase the coefficient $\alpha$
in mesoscopic experiments:
one is to increase the electron density $n_s$ in QPC
since $\alpha$ is proportional to $n_s$;
and the other is to vary the gate bias which
controls the magnitude of the SO-interaction.
On the other hand, a larger effective mass of charge
carriers would also be helpful for observing the SO-induced oscillations in experiments.
Secondly, the SO coupling in the leads has been neglected 
in the present study. This approximation
may enhance the oscillations observed above although the Rashba SO coupling in QPC
is larger than that in the leads.\cite{Qpcnote} 
It is extremely difficult to include the SO coupling in the leads using the present method because
the analytical  wave function in the leads should be given;
while it  appears to be unsolvable in the leads with the SO coupling .
Nevertheless, with the advancement of nanotechnology, it is possible to  
fabricate a mesoscopic structure where the SO coupling is important  only in the QPC 
(i.e., negligible in the two leads). In such systems, it could be easier to observe the SO-induced effects
 predicted here. 

In conclusion, we have generalized a recursive Green's function
approach to calculate numerically
the spin-dependent conductance
in mesoscopic structures in the presence of SO-interaction
and magnetic field. The effect of SO-interaction
on the conductance of a QPC is studied in detail.
Some interesting oscillations in conductance
induced by the SO-interaction have been observed
numerically, which may be tested
by future mesoscopic experiments.

\acknowledgements{We thank Prof. W. Y. Zhang for
his helpful discussions. This work was supported by a RGC grant of Hong Kong
under grant number HKU7118/00P.}

\appendix
\section{}
In the appendix we present the recursive Green's function approach to calculate
the spin-dependent transmission $t_{lj}^{\sigma\sigma'}$.
The electron wave functions in the strip $n$
of the leads can be expressed as a linear
combination of eigenfunctions of a straight infinite lead at a given
energy $E$.
If the incident electron from the left lead is in the
$ (l,\sigma)$ channel with $l$ and $\sigma$ as the subband and spin
indices,
the scattering states are represented as
\begin{eqnarray}
\label{left-rightwave}
&|& n)_{left} = e^{ik_l n}\phi_{k_l\sigma}
+\sum\limits_{j\sigma'} r_{lj}^{\sigma\sigma'}
e^{-ik_jn}\phi_{k_j\sigma'}, \\
&|& n)_{right} = \sum\limits_{j\sigma'} t_{lj}^{\sigma\sigma'}
e^{ik_jn}\phi_{k_j\sigma'},
\end{eqnarray}
where $t_{lj}^{\sigma,\sigma'}$
($r_{lj}^{\sigma,\sigma'}$)
is the transmission (reflection) amplitudes 
for the incident channel $(l,\sigma)$
and out-going channel $(j,\sigma')$,
and the wave functions $\phi_{k_j\sigma}$
are given by
$ \phi_{k_j\sigma}=
\phi_{k_j}\otimes \chi(\sigma_z)$ with
$$
\phi_{k_j}=\sqrt{\frac{2}{M+1}}(sin\frac{\pi j}{M+1},\cdots,sin\frac{\pi j m}{M+1},
\cdots, sin\frac{\pi j M}{M+1})^{T_r},
$$
and
$$
\chi(+)=
\left (
\begin{array}{l}
1\\
0 
\end{array}
\right ),\ \ \chi(-)=\left (
\begin{array}{l}
0\\
1
\end{array}
\right ).
$$
Here the sign $T_r$ denotes the transposition of matrix.
The number of open channels and the value of
$k_j$ are determined by the Fermi energy from the
energy dispersion in ideal leads 
$$
E=4t-2t(cosk_j+cos\frac{\pi j}{M+1}).
$$
Only if $k_j$ is real do we have a propagating state;
we have an evanescent state when $k_j$ is imaginary.
All real $k_j$'s are positive, because we
have already considered separately the states propagating to the right
(incident and transmitted wave) or to the left (reflected wave).
Following the recursive Green's function technique, 
we find
\begin{equation}
\label{coefficient}
t_{lj}^{\sigma\sigma'}
=\sqrt{v_j/v_l}\phi^+_{k_j\sigma'}
G_{N+1,0}\Theta(\sigma) \phi_{k_l\sigma}
e^{-ik_j L_x},
\end{equation}
where
$$
v_j=\frac{1}{\hbar}\frac{\partial E}{\partial k_j},
$$
$$
\Theta(\sigma)=-2it\sum\limits_{l=1}^{M}sink_l
[Q_l\otimes \zeta (\sigma_z)],
$$
with
$$
(Q_l)_{pp'}=\frac{2}{M+1}sin\frac{l\pi p}{M+1}sin\frac{l\pi p'}{M+1},
\ (p,p'=1,\cdots,M),
$$
$$
\zeta (+)
=\left (
\begin{array}{lc}
1 & 0\\
0 & 0
\end{array}
\right ),\ \ \zeta (-)=\left (
\begin{array}{lc}
0 & 0\\
0 & 1
\end{array}
\right ).
$$
The set of vectors $\phi^+_{k_j\sigma'}$
are the duals of the set $\phi_{k_l\sigma}$, defined by
$\phi^+_{k_j\sigma'} \phi_{k_l\sigma}
=\delta_{jl}\delta_{\sigma'\sigma}$.
$G_{N+1,0}$ is the retarded Green's function for
the scattering region between two ideal
leads, which can be obtained by a set of
recursion formulas in a matrix form,~\cite{greenfunction}
\begin{eqnarray}
\label{Gfunction}
G_{n'+1,0} &=& g^{n'+1}H_{n'+1,n'}G_{n',0},\ \ (0\leq n' \leq N) \\
g^{n'+1} &=& [E-\widetilde{H}_{n'+1}-\Lambda^{n'}]^{-1},  \\
(\Lambda^{n'})_{pp'} &=&  e^{i2\pi(p-p')\beta}V_x^+ (g^{n'})_{pp'}V_x,
\ \ (p,p'=1,\cdots,M)
\end{eqnarray}
by iteration starting form
$g^0=G_{0,0}=(E-\widetilde{H}_0)^{-1}$, where
$\widetilde{H}_0=H_0-tF(\sigma_{in})$,
$\widetilde{H}_l=H_l\ (1\leq l \leq N)$,
$\widetilde{H}_{N+1}=H_{N+1}-tF(\sigma_{out})$
with $\sigma_{in}$ $(\sigma_{out})$ as the spin state of the incident
(outgoing) electrons, and
\begin{equation}
\label{F_sigma}
F(\sigma)=\sum\limits_{j=1}^{M} e^{ik_j}[Q_j\otimes \zeta(\sigma_z)],
\end{equation}
It is worth pointing out that
each element of Green's function in the above
recursion formula is given by a quaternion number
in the presence of the SO-interaction or the Zeeman effect.\cite{Complex}
In the numerical calculations of
$G_{N+1,0}$ and $\Theta(\sigma)$ in Eq.(\ref{coefficient}),
we have used explicitly 
the multiplication table
for quaternion number.


\begin{references}

\bibitem{Meir} Y. Meir, Y. Gefen, and O. Entin-Wohlman,
Phys. Rev. Lett. {\bf 63}, 798 (1989).
\bibitem{Loss} D. Loss and P. M. Goldbart, Phys. Rev. B {\bf 45},
13 544 (1992).
\bibitem{Aronov} A. G. Aronov and Y. B. Lyanda-Geller,
Phys. Rev. Lett. {\bf 70}, 343 (1993).
\bibitem{Zhou} Y. C. Zhou, H. Z. Li, and X. Xue,
Phys. Rev. B {\bf 49}, 14 010 (1994);
S. L. Zhu, Y. C. Zhou, and H. Z. Li, {\sl ibid.}
{\bf 52}, 7814 (1995). Z. D. Wang and S. L. Zhu,
{\sl ibid.} {\bf 60}, 10 668 (1999).
\bibitem{Zhu} S. L. Zhu and Z. D. Wang, Phys. Rev. Lett.
{\bf 85}, 1076 (2000);
S. L. Zhu, Z. D. Wang, and Y. D. Zhang, Phys. Rev. B
{\bf 61}, 1142 (2000).
\bibitem{Morpurgo} A. F. Morpurgo, J. P. Heida, T. M. Klapwijk,
B. J. van Wees, and G. Borghs, Phys. Rev. Lett. {\bf 80}, 1050 (1998).
\bibitem{Moroz} A. V. Moroz and C. H. W. Barnes, Phys. Rev. B
{\bf 60}, 14 272 (1999);
{\sl ibid.} {\bf 61}, R2464 (2000).
\bibitem{Bulgakov} E. N. Bulgakov, K. N. Pichugin, A. F. Sadreev,
P. St\u{r}eda, and P. \u{S}eba, Phys. Rev. Lett. {\bf 83}, 376 (1999).
\bibitem{Rashba} In a system of two-dimensional electron gas (2DEG)
in the absence of external magnetic field, the
spin of a moving electron 
feels an effective magnetic field
induced by a perpendicular interface electric field. 
This kind of SO-interaction is 
commonly referred to as the Rashba SO-interaction.
See E. I. Rashba, Sov. Phys. Solid State {\bf 2}, 1109 (1960).
\bibitem{Luo} J. Luo, H. Munekata, F. F. Fang, and P. J. Stiles,
Phys. Rev. B {\bf 38}, 10 142 (1988);
J. Nitta, T. Akazaki, H. Takayanagi, and T. Enoki,
Phys. Rev. Lett. {\bf 78}, 1335 (1997);
J. P. Heida, B. J. van Wees, J. J. Kuipers, T. M. Klapwijk, and G. Borghs,
Phys. Rev. B {\bf 57}, 11 911 (1998).
\bibitem{Lommer} G. Lommer, F. Malcher, and U. R\"{o}ssler,
Phys. Rev. Lett. {\bf 60}, 728 (1988).
\bibitem{Wees} B. J. van Wees, H. van Houten, C. W. Beenakker,
J. G. Williamson, L. P. Kouwenhoven, D. van der Marel, and
C. T. Foxon. Phys. Rev. Lett. {\bf 60},
848 (1988); B. J. van Wees, L. P. Kouwenhoven, E. M. M. Willems, C. J. P. M. Harmans,
J. E. Mooij, H. van Houten, C. W. J. Beenakker,
J. G. Williamson, and C. T. Foxon, Phys. Rev. B {\bf 43}, 12 431 (1991).
\bibitem{Wharam} D. A. Wharam, T. J. Thornton, R. Newbury, M. Pepper,
H. Ahmed, J. E. F. Frost, D. G. Hasko, D. C. Peacock, D. A. Ritchie, and
G. A. C. Jones, J. Phys. {\bf B 21}, L209 (1988).
\bibitem{Thomas} K. J. Thomas, J. T. Nicholls, M. Y. Simmons,
M.Pepper, D. R. Mace, and D. A. Ritchie, Phys. Rev. Lett. {\bf 77},
135 (1996).
\bibitem{Ando} T. Ando, Phys. Rev. B {\bf 44}, 8017 (1991).
\bibitem{Sanvito} S. Sanvito, C. J. Lambert, J. H. Jefferson,
and A. M. Bratkovsky, Phys. Rev. B {\bf 59}, 11 936 (1999).
\bibitem{Cuevas} J. C. Cuevas, A. L. Yeyati, and A. Mart\'{l}n-Rodero,
Phys. Rev. Lett. {\bf 80}, 1066 (1998); J. C. Cuevas, A. Mat\'{l}n-Rodero, and A. L. Yeyati,
Phys. Rev. B {\bf 54}, 7366 (1996).
\bibitem{Datta2} S. Datta, Electronic transport in mesoscopic
systems (the Press Syndicate of the Cambridge University, New York, 1995),
pp145-146.
\bibitem{Lee} P. A. Lee and D. S. Fisher, Phys. Rev. Lett. {\bf 47},
882 (1981); A. MacKinnon, Z. Phys. B {\bf 59}, 385 (1985).
\bibitem{Landauer} R. Landauer, IBM J. Res. Dev. {\bf 1}, 223 (1957);
M. B\"{u}ttiker, Phys. Rev. Lett. {\bf 57}, 1761 (1986).
\bibitem{Buttiker} M. B\"{u}ttiker, Phys. Rev. B {\bf 41}, 7906 (1990).
\bibitem{Sizes} For different system sizes,
we observe that the main properties 
are the same as long as $N,\ M\geq 5$, which implies that the
transport properties addressed below are
almost independent on the system size.
\bibitem{small} Equation (\ref{reflection}) is clearly oversimplified 
since $R_{j\sigma}$ is assumed to be the reflection probability
on the interfaces between the ideal leads and the scattering region,
we can nevertheless apply it to understand some features of 
the transmission oscillations, as done by van Wess {\sl et al} in explaining
the oscillations observed in their experimental data.\cite{Wees}
\bibitem{greenfunction} The results in spin-independent cases can be
found in Refs.\cite{Ando,Lee}.
\bibitem{Complex} Each element in the above Green's function
is a complex number if only the Peierls phase
factor associated with the magnetic field is considered
(both the Zeeman effect and the SO coupling are ignored).
\bibitem{Qpcnote}  The QPC is often produced by a negative gate while
the negative gate enhances the Rashba SO coupling. This effect
was verified experimentally by Nitta {et al} in Ref.\cite{Luo}.

\end{references}
\end{document}